\documentclass[fleqn,10pt]{wlscirep}
\title{Multiple imputation using dimension reduction techniques for high-dimensional data}
\usepackage{bm}
\usepackage{setspace}

\author[1]{Domonique W. Hodge}
\author[2]{Sandra E. Safo}
\author[3,*]{Qi Long}
\affil[1]{Department of Biostatistics and Bioinformatics, Emory University, USA}
\affil[2]{Division of Biostatistics, University of Minnesota, USA}
\affil[3]{Department of Biostatistics, Epidemiology and Informatics, University of Pennsylvania, USA}

\begin{abstract}
	Missing data present challenges in data analysis.  Naive analyses such as complete-case and available-case analysis may introduce bias and loss of efficiency, and produce unreliable results. Multiple imputation (MI) is one of the most widely used methods for handling missing data which can be partly attributed to its ease of use.
	However, existing MI methods implemented in most statistical software are not applicable to or do not perform well in high-dimensional settings where the number of predictors is large relative to the sample size. To remedy this issue, we develop an MI approach that uses dimension reduction techniques. Specifically, in constructing imputation models in the presence of high-dimensional data our approach uses sure independent screening followed by either sparse principal component analysis (sPCA) or sufficient dimension reduction (SDR) techniques. Our simulation studies, conducted for high-dimensional data, demonstrate that using SIS followed by sPCA to perform MI achieves better performance than the other imputation methods including several existing imputation approaches. We apply our approach to analysis of gene expression data from a prostate cancer study.
	
\end{abstract}

\begin{document}

\flushbottom
\maketitle
\newpage

\doublespacing

\section*{Introduction}

Appropriate handling of missing data requires an understanding of its source and structure. It is well known that naive analyses such as complete-case and available-case analysis may introduce bias and loss of efficiency, and produce unreliable results.  Multiple imputation (MI) \cite{Rubin1976, rubin1987multiple} is one of the most widely used methods which can be partly attributed to its ease of use. The basic idea underlying MI is to replace missing values $M$ times by "plausible values" drawn from their posterior predictive distributions given the observed data. Multiply imputed data sets are generated to account for sampling variability and uncertainty of imputing missing values. Then each data set completed by imputation is analyzed using the standard complete-data methods and the estimates obtain from these analyses are combined  using Rubin's rule \cite{rubin1987multiple} to create one statistical inference summary. A key advantage of MI is that the imputation model can be operationally distinct from the subsequent analyses (target analysis that would be performed in the absence of missing data). The use of MI has been investigated in various settings and detailed reviews are provided elsewhere. \cite{harel2007multiple, carpenter2012multiple}

\subsection*{The Problem}
The validity of MI is predicated on several assumptions. First, the missing at random (MAR) \cite{little2014statistical} mechanism is often assumed and implies the missingness is not associated with the missing values conditional on observed data.\cite{Rubin1976} Our current work assumes that the incomplete data are MAR. Second, Meng \cite{meng1994multiple} suggested that the imputation model be congenial or general enough to preserve any associations among variables that may be the target in the imputed data analyses. Furthermore, a general imputation model that is close to the true model allows for accommodation of a wide range of statistical models that can be used on the imputed data sets.
In order to construct a reasonable, general imputation model, a major issue is to not exclude any important predictors, since excluding important variables may lead to imputation models that are not as general than the subsequent analysis and will potentially bias the results. However, in practice it is not feasible to specify all possible relevant predictors and their interactions in an imputation model. A more challenging problem arises in the presence of high-dimensional data where the number of variables is larger than or approximately equal to the sample size. Largely due to the advancement of technology, the amount of data collected is rapidly increasing which give rise to high-dimensional data. Examples of high-dimensional data include genomics, proteomics and functional magnetic resonance imaging data. These data often contain missing values, yet there has been limited work in developing approaches for handling missing data in the presence of high-dimensional data. Standard MI approaches implemented in most statistical software perform poorly or fail in the presence of high-dimensional data.\cite{Zhao2013} %When conducting imputation in such settings, MI requires model trimming to construct imputation models.

\subsection*{Existing approaches for MI in the Presence of High-Dimensional Data}
Model trimming is essential to construct imputation models in the presence of high-dimensional data. Stekhoven et al. \cite{Stekhoven2012} used a classification technique, namely random forest (RF), to impute missing values in high-dimensional data.  The variable with missing values is treated as the response variable and other (auxiliary) variables are used for bootstrap aggregation of multiple regression trees to potentially reduce overfitting. The predictions are combined from trees to improve accuracy of prediction of the missing values.  However, the selection of tuning parameters such as the number of trees and number of nodes needs further investigation.  Liao et al. \cite{liao2014missing} proposed another imputation approach for high-dimensional data  which is a variation of a $K$-nearest-neighbor imputation. For a missing value, the method seeks its $K$ nearest variables (KNN\_V) or subjects (KNN\_S) and imputes by a weighted average of observed values of the similar neighbors. Although the method was shown to perform well in their simulations where the performance was evaluated based on comparisons between true and imputed values, it does not properly propagate the uncertainty in estimating the parameters in the imputation model and hence it is not proper in the sense of Rubin\cite{rubin1987multiple}. Improper imputation can lead to biased inference in the subsequent analyses.

Zhao and Long\cite{Zhao2013} proposed an MI approach for high-dimensional data based on regularized regression that does account for the uncertainty in imputation.  Specifically, they investigated the use of MI through direct and indirect use of regularized regression. In the former, regularized regression is used for both variable selection and parameter estimation for imputation models; in the latter regularized regression is only used for model trimming.  Direct use of regularized regression in MI was shown to achieve superior performance in the settings considered in their work.  They also proposed an MI method using the Bayesian lasso (Blasso)\cite{park2008bayesian} to estimate and select important variables in imputation models. However, these methods also have some limitations and particularly they may not yield good performance when the true imputation model is large. To tackle this challenge, we consider an alternative approach to constructing imputation models by incorporating dimension reduction techniques.

\subsection*{Dimension Reduction Techniques for High-Dimensional Data}
Screening is an effective strategy to deal with high dimensionality. In particular, sure independent screening (SIS) \cite{fan2008sure} is a method which is based on correlation learning which filters out the features that have weak correlation with the response.  Another dimension reduction technique is sparse principal component analysis (sPCA).\cite{Zou2006} The commonly used principal component analysis (PCA) seeks linear combinations of $p$ variables such that the derived components capture maximum variance. Yet, a drawback of PCA is that the loadings of all $p$ variables are typically nonzero, which is often hard to interpret. Zou et al.\cite{Zou2006} modified PCA by using the lasso penalty (sPCA\_ST) to shrink some loadings to zero, allowing for identification of important features. More recently, other authors proposed adjustments to sparse principal component analysis. Witten et al.\cite{Witten2009} used penalized matrix decomposition (sPCA\_PMD), a regularized version of the singular value decomposition, to create sparse loadings. Lee et al. \cite{lee2012sparse} proposed two approaches to modify sPCA by using the lasso \cite{tibshirani1996regression} (sPCA\_L)   and adaptive lasso \cite{zou2006adaptive} (sPCA\_AL)  penalty terms.

Alternatively, we can use sufficient dimension reduction regression \cite{drpack} (SDR) to find relevant predictors in imputation models.  SDR seeks to find $d$ linearly independent linear combinations such that all the information about the regression is contained in the $d$ linear combinations and $d$ is typically considerably less than the number of variables (namely, $p$). There are several variations of SDR including sliced inverse regression (SIR) \cite{li1991sliced}, sliced average variance estimates (SAVE) \cite{dennis2000save}, and principal Hessian directions (PHD) \cite{li1992principal}.

In this paper, we propose a new MI approach that imputes the missing values by first screening for relevant predictors of a variable with missing values.  Using the screened variables, we further reduce dimensions by applying SDR or sPCA and use the resulting linear combinations to construct imputation models. The remainder of this paper is organized as follows. In the next section, we describe the proposed approach based on SPCA and SDR. In the following section, we perform simulations to evaluate the performance of the proposed approach in comparison with several existing approaches including Blasso, DAlasso, KNN\_S, KNN\_V, and RF in the presence of high-dimensional data. In the fourth section we illustrate the new proposed approach using genomics data from a prostate cancer study. We conclude with a discussion in the last section.

\section*{Methodology}\label{meth}
To fix ideas, let $\bm{Y}$ denote a set of $p$ variables observed for a random sample of $n$ observations. Denote by $\bm{Y}_{obs}$ the observed components of $\bm{Y}$ and by $\bm{Y}_{mis}$ the missing components of $\bm{Y}$. Suppose that $\bm{Y}=(\bm{Y}_{obs},\bm{Y}_{mis})$ follows a model $\pi(\bm{Y}\vert \bm{\beta})$ where $\bm{\beta}$ is a set of parameters and the missing data mechanism is missing data at random (MAR). Under ignorability, the standard imputation framework can be represented by \eqref{eq1}
\begin{eqnarray}\label{eq1}
\pi(\bm{Y}_{mis}\vert \bm{Y}_{obs})=\int \pi(\bm{Y}_{mis}\vert \bm{Y}_{obs},\bm{\beta})\pi(\bm{\beta}\vert \bm{Y}_{obs})d\bm{Y}.
\end{eqnarray}
Specifically, one can first generate a random draw from the posterior distribution of $\bm{\beta}$
\begin{eqnarray*}
\bm{\beta}^{(m)}\sim \pi(\bm{\beta}\vert \bm{Y}_{obs}),
\end{eqnarray*}
and then generate a random draw of the missing values from their posterior predictive distributions
\begin{eqnarray*}
\bm{Y}_{mis}^{(m)}\sim \pi(\bm{Y}_{mis}\vert \bm{Y}_{obs},\bm{\beta}^{(m)}),
\end{eqnarray*}
where $m=1,...,M$ and $M$ is the number of imputed data sets. \cite{Schafer1999}

%One imputation is created by a random draw from the predictive distribution (\ref{eq2}) and multiply imputed data sets can be generated by $M$ draws from the predictive distribution (\ref{eq3}).

\indent For ease of exposition, we describe our proposed approach in a setting where only one variable $\bm{y}_1$ has missing values with the remaining variables $\{\bm{y}_2,...,\bm{y}_p\}$ fully observed, and all $\bm{y}$ are continuous variables. Let $n_1$ denote the number of complete cases with all variables observed and $n_2$ the number of incomplete cases with $y_1$ missing ($n=n_1+n_2$). Define $\bm{y}_{obs,1}=(y_{1,1},y_{2,1},...,y_{n_1,1})^T	$ as the first $n_1$ observed components of $\bm{y}_1$ and its complement as $\bm{Y}_{obs,-1}=(\bm{y}_{1,-1},\bm{y}_{2,-1},...,\bm{y}_{n_1,-1})^T$ with $\bm{y}_{i,-1}=(y_{i,2},y_{i,3},...,y_{i,p})$, which together form the set of complete cases.  Define $\bm{y}_{mis,1}=(y_{n_1+1,1},y_{n_1+2,1},...,y_{n,1})$ as the $n-n_1$ missing components of $\bm{y}_1$ and its complement as $\bm{Y}_{mis,-1}=(\bm{y}_{n_1+1,-1},\bm{y}_{n_1+2,-1},...,\bm{y}_{n,-1})$, which together form the set of incomplete cases. Of note, $\bm{Y}_{mis,-1}$ is observed. It follows that the observed data are $(\bm{y}_{obs,1},\bm{Y}_{obs,-1},\bm{Y}_{mis,-1})$ and the missing data are $\bm{y}_{mis,1}$. The imputation model \eqref{eq1} reduces to
\begin{eqnarray}\label{eq1a}
\pi(\bm{y}_{mis,1}\vert \bm{y}_{obs,1},\bm{Y}_{obs,-1},\bm{Y}_{mis,-1})=\int \pi(\bm{y}_{mis,1}\vert \bm{Y}_{mis,-1},\bm{\beta})\pi(\bm{\beta}\vert \bm{y}_{obs,1},\bm{Y}_{obs,-1})d\bm{\beta}.
\end{eqnarray}
To complete the imputation model \eqref{eq1}, we can posit a regression model with $\bm{y}_1$ as the outcome
\begin{eqnarray}\label{lin}
\bm{y}_1=\delta_0+\bm{Y}_{obs,-1}\bm{\delta}+\epsilon
\end{eqnarray}
where $\epsilon \sim N(\bm{0},\sigma^2 I_{n_1})$ and $\bm{\beta}=(\delta_0,\bm{\delta},\sigma^2)^T$.  Model \eqref{lin} can be fitted using the set of complete cases. However, when $p\gg n$, standard regression techniques such as ordinary least squares fail and it is imperative to perform variable selection or dimension reduction when fitting model \eqref{lin}. As demonstrated in our simulations, when the true model for  \eqref{lin} is large, i.e., the number of important predictors in \eqref{lin} is large relative to $n$, imputation methods based on regularized regression may yield unsatisfactory performance.

We propose to use dimension reduction techniques when constructing imputation models, specifically, applying SIS followed by either sPCA or SDR before fitting model \eqref{lin}. The proposed imputation approach is detailed as follows:

\begin{enumerate}
	\item In the first step, SIS is performed using the complete cases to find a subset of $v$ variables that are predictive of the incomplete variable $y_1$. Let $\{t_1,...,t_v\}$ index the subset of $v$ variables selected from $y_2,\ldots,y_p$ using SIS, where $v<p-1$.
	\item In the second step, we achieve further dimension reduction via either sPCA or SDR, as $v$ can be still large relative to $n$.
	\begin{enumerate}
		\item Applying sPCA to the set of $v$ variables selected in the first step and using all $n$ observations, we obtain
		\begin{eqnarray*}
			\bm{z}_1 &=& \alpha_{1,1}\bm{y}_{t_1}+\alpha_{1,2}\bm{y}_{t_2}+...+\alpha_{1,v}\bm{y}_{t_v}\\
			\bm{z}_2 &=& \alpha_{2,1}\bm{y}_{t_1}+\alpha_{2,2}\bm{y}_{t_2}+...+\alpha_{2,v}\bm{y}_{t_v}\\
			\vdots \\
			\bm{z}_{v}&=& \alpha_{v,1}\bm{y}_{t_1}+\alpha_{v,2}\bm{y}_{t_2}+...+\alpha_{v,v}\bm{y}_{t_v}
		\end{eqnarray*}
		where the linear combinations $\bm{z}_1,\bm{z}_2,...,\bm{z}_{v}$ are the principal components, and $\bm{\alpha}_1,\bm{\alpha}_2,...,\bm{\alpha}_{v}$ are the loading vectors.  We select $\bm{z}_1$, $\bm{z}_2$, ... , $\bm{z}_d$ in sPCA by either choosing the first principal component or the first $d$ principal components that explain at least 60 or 80\% of the total variance. It is important to note that $d$ is typically substantially less than the original number of variables $p$ and the sample size $n$.
		
		\item Alternatively, applying SDR to the set of $v$ variables selected in the first step, we obtain
		\begin{eqnarray*}
			\bm{z}_1 &=& \gamma_{1,1}\bm{y}_{t_1}+\gamma_{1,2}\bm{y}_{t_2}+...+\gamma_{1,v}\bm{y}_{t_v}\\
			\bm{z}_2 &=& \gamma_{2,1}\bm{y}_{t_1}+\gamma_{2,2}\bm{y}_{t_2}+...+\gamma_{2,v}\bm{y}_{t_v}\\
			\vdots \\
			\bm{z}_{d}&=& \gamma_{d,1}\bm{y}_{t_1}+\gamma_{d,2}\bm{y}_{t_2}+...+\gamma_{d,v}\bm{y}_{t_v}
		\end{eqnarray*}
		where $\bm{y}_1$ is used as the response variable, $\gamma$'s are the estimated coefficients in SDR using the set of complete cases, and $d$ ($d<v$) is chosen by using an asymptotic test for PHD and permutation tests for SAVE and SIR \cite{drpack}. The most notable difference between sPCA and SDR is that SDR uses the variable with missing values, namely $\bm{y}_1$, as the outcome to guide dimension reduction.
	\end{enumerate}
	
	After obtaining $\bm{z}_1$, $\bm{z}_2$, ... , $\bm{z}_d$ via either sPCA or SDR,  define $\bm{Z}_{obs}=(\bm{z}_{1},\bm{z}_{2},...,\bm{z}_{n_1})^T$ for the complete cases and $\bm{Z}_{mis}=(\bm{z}_{n_1+1},\bm{z}_{n_1+2},...,\bm{z}_{n})$ for the incomplete cases. Of note, both $\bm{Z}_{obs}$ and $\bm{Z}_{mis}$ can be calculated, since they involve only a subset of $y_2,\ldots,y_p$.
	
	\item In the third step, we replace $\bm{Y}_{obs,-1}$ and $\bm{Y}_{mis,-1}$ with $\bm{Z}_{obs}$ and $\bm{Z}_{mis}$ respectively in \eqref{eq1a} and \eqref{lin} and conduct imputation accordingly. More specifically, using model \eqref{lin} with $\bm{y}_{obs,1}$ as the outcome variable and $\bm{Z}_{obs}$ as predictors, we randomly draw $\hat{\bm{\beta}}^{(m)}$ from its posterior distribution and then impute $\bm{y}_{mis,1}$ using $\bm{Z}_{mis}$ by drawing randomly from the conditional posterior predictive distribution $\pi(\bm{y}_{mis,1}\vert \bm{Z}_{mis},\hat{\beta}^{(m)})$, where $m=1,\ldots,M$ for $M$ imputations.
\end{enumerate}

Once the missing data are multiply imputed, subsequent analyses such as multiple regression or logistic regression are performed for each of the $M$ imputed datasets. Analysis results are then combined for statistical inference using Rubin's combining rule \cite{rubin1987multiple}.

\section*{Simulation studies}\label{sim}
We conduct simulation studies to evaluate the performance of our proposed approach. In the second step of our proposed approach, we use either sPCA or SDR. For sPCA, we consider four variations, namely, sPCA\_ST \cite{zou2006adaptive}, sPCA\_PMD \cite{Witten2009}, sPCA\_L \cite{lee2012sparse}, and sPCA\_AL \cite{lee2012sparse}. The sPCA\_ST and sPCA\_PMD methods are both implemented in  R package \textbf{PMA}\cite{PMApack}. The sPCA\_L and sPCA\_AL methods are both implemented in R code provided on the authors website.\cite{lee2012sparse} For SDR, we consider three variations, namely, sliced average variance estimates \cite{dennis2000save} (SDR\_SAVE), sliced inverse regression \cite{li1991sliced} (SDR\_SIR), and principal Hessian directions \cite{li1992principal} (SDR\_PHD). The SDR methods are implemented in the R package \textbf{dr}\cite{drpack}. We compare our approach to several existing imputation methods proposed by Zhao and Long \cite{Zhao2013}, Liao et al.\cite{liao2014missing}, and Stekhoven et al.\cite{Stekhoven2012}. Zhao and Long \cite{Zhao2013} used Bayesian lasso regression (Blasso) and adaptive lasso with direct use of regularized regression (DAlasso) to conduct MI in high-dimensional data.  Liao et al. \cite{liao2014missing}  used variations of k-nearest neighbors, namely,  KNN\_S and KNN\_V for imputation of missing values. Stekhoven et al.\cite{Stekhoven2012} proposed random forest (RF) for MI.  We also include the standard parametric MI procedure implemented in the R package \textbf{mice}\cite{micepack} with the default method of Bayesian linear regression.

In our simulations we focus on estimating the regression coefficients $\hat{\theta}$ from linear regression in the presence of missing data. We use $\hat{\theta}$ obtained from the imputed data sets to compare the performance across different imputation methods. To have a point of reference for bias and efficiency for estimating $\theta$, we apply a gold standard (GS) method that estimates $\theta$ using the underlying complete data before missing data are generated. We also perform a complete-case analysis (CC), in which only the set of complete cases are used in data analysis.

\subsection*{Simulation setup}
We vary several factors including the total number of variables ($p$), the number of variables in the true imputation model ($c$), and the correlation among the data ($\rho$). Simulations  are  carried out with 500 Monte Carlo (MC) datasets and the sample size is fixed at $n = 100$ in each MC dataset. Each simulated data set includes the fully observed outcome variable ($\bm{w}$) and the set of predictors and auxiliary variables $\bm{Y}=(\bm{y}_1,\bm{y}_2,...,\bm{y}_p)$ .  The variable $\bm{y_1}$ contains missing values $\bm{y}_{mis,1}=(y_{n_1+1,1},y_{n_1+2,1},...,y_{n,1})$.   The details of the simulation set-up are as follows:
\begin{itemize}
	\item[I.] $\bm{Y_{-1}}=(\bm{y}_2,\bm{y}_3,...,\bm{y}_p)$ is generated from a multivariate normal distribution with mean $(0,0,...,0)_{p-1}$ and a first-order autoregressive covariance matrix with autocorrelation, denoted by $\rho$, varying as 0.1, 0.5, or 0.9. We also consider a block diagonal covariance matrix having main diagonal blocks square matrices with the off-diagonal blocks as zero matrices. The main diagonal block matrices are composed of compound symmetric matrices with $\sigma^2=2$ on the diagonals and $\sigma^2\rho$ on the off-diagonals, where $\rho$ is again varied as 0.1, 0.5, and 0.9. We consider settings with $p=200$ and $p=1000$.		
	\item[II.] For each combination of $p$ and $\rho$, $\bm{y}_1$ is generated from a normal distribution such that
	$y_1 \sim \mbox{N}(\bm{Y}_T\bm{\eta},1)$, where $T$ represents the set of the variables in the true imputation model with a cardinality of $c$. Two cases are considered with the corresponding design matrices $\bm{Y}_T=(\bm{y}_2,\bm{y}_3,\bm{y}_{50},\bm{y}_{51}),$ $(\bm{y}_2,...,\bm{y}_{51},\bm{y}_{100},...,\bm{y}_{149})$ for $c = 4$ and $c=100$, respectively.  We set $\bm{\eta}=\bm{1}'\times 1,\bm{1}'\times 0.05$  to include the intercept for $c=4$ and $c=100$, respectively. The values of $\bm{\eta}$ are chosen to fix the signal-to-noise ratio when generating $y_1$. Of note, $c=100$ corresponds to the case where the size of the true imputation model is large relative to the sample size.
	\item[III.] Given $\bm{Y}, \bm{w}$ is generated from a normal distribution $w\sim\mbox{N}(\theta_0+\theta_1y_1+\theta_2y_2+\theta_3y_{10},\sigma_2=3),$ where all $\theta=1.$		 
	\item[IV.] The response indicator $R$ for $\bm{y}_1$, which is 1 if $\bm{y}_1$ is observed and 0 otherwise, is generated from a logistic model, $\mbox{logit}[\mbox{Pr}(R=1\vert \bm{Y}_{-1},\bm{w})]=-1-0.1\bm{y}_2+2\bm{y}_3-10\bm{w}$ which results in an average of 31\% of $\bm{y}_1$ missing.
\end{itemize}

We conduct imputation of $y_{mis,1}$ using each imputation approach considered. The subsequent analysis for the imputed data is performed using a linear regression of $\bm{w}$ on the imputed $\bm{y}_1$ and fully observed $\bm{y}_2$ and $\bm{y}_{10}$.  For our proposed method and the existing MI based methods,  we use the R package \textbf{mice} to multiply impute the missing data with its default method of Bayesian linear regression with a ridge parameter value of $10^-5$. Thirty imputed data sets are generated for each MI method to use in subsequent analyses. Rubin's rule \cite{rubin1987multiple} is used to pool the estimates to obtain $\hat{\bm{\theta}}$ and their standard errors.

\subsection*{Results}

The simulation results are summarized for $\hat{\theta}_1$ which is the parameter estimate that is associated with the incomplete variable $\bm{y}_1.$ Tables 1 to 4 present the mean bias of $\hat{\theta}_1$ (Bias), mean standard error of $\hat{\theta}_1$ (SE), Monte Carlo standard deviation of $\hat{\theta}_1$ (SD), mean square error of $\hat{\theta}_1$ (MSE), and coverage rate of the 95\% confidence interval of $\hat{\theta}_1$ (CR). In addition to comparing different methods, we evaluate the effect of the dimension of the data ($p$),  the number of variables in the true imputation model ($c$), and the correlation among the data $\bm{Y_{-1}}$ ($\rho$). Within each table, $p$ and $\rho$ are varied while $c$ is fixed. In both Tables 1 and 2, the covariance structure is an autoregressive matrix. More specifically,  $c = 4$ in Table 1 and $c = 100$ in Table 2. In Table 3 and Table 4, the covariance structure is a block diagonal matrix with compound symmetric blocks. We let $c=4$ in Table 3 and let $c = 100$ in Table 4. In our proposed approach, we use the R package \textbf{SIS} \cite{SISpack} with the default vanilla method to find a subset of $v$ variables that are predictive of the incomplete variable $y_1$. In each setting, SIS selects between 10 and 17 variables. SPCA is conducted using four variations and the number of principal components (PCs) are selected by either choosing the first principal component or the first $d$ principal components that explain at least 60 or 80\% of the total variance. However, we observe superior performance using one principal component; as a result, we only present the results based on one principal component.  

%%%% comparing our approach to the existing approaches
In comparing the existing methods to our new proposed methods, we observe that our new proposed method outperforms the existing methods. When the size of the true imputation model is small ($c=4$) relative to $n$, the Blasso imputation method of Zhao and Long \cite{Zhao2013} yields modest bias. In contrast, the CC, MI, DAlasso, KNN and RF methods, in general, yield substantial bias and inadequate CRs. More importantly, our methods outperform all the existing methods, including Blasso, in terms of bias, MSE, and coverage rates. Furthermore, as $c$ increases to 100, there is considerably more pronounced deterioration in the performance of Blasso compared to our proposed approach.  Our proposed method has minimal bias, coverage rates near the nominal level, and overall superior performance as compared to the existing methods, irrespective of whether the true imputation model is small or large relative to $n$.

%%%% comparing among our approach	
Among the two proposed methods, sPCA generally achieves better performance than SDR. When the size of the true active set in the imputation model is small, that is, $c = 4$ (Table 1), all sPCA and SDR variations exhibit negligible bias and coverage near the nominal level.  Within SDR methods, SDR\_SIR tends to achieve slightly better performance than SDR\_SAVE  and SDR\_PHD in terms of bias, MSE, and coverage. When the size of the true active set is large ($c=100$), the improved performance of sPCA compared to SDR is more pronounced when correlation among the data is small ($\rho = 0.1$) or moderate  ($\rho = 0.5$). As the number of variables in the true imputation model increases, the performance of the SDR methods slightly deteriorates. While SDR achieves satisfactory performance in terms of bias and MSE when $c$ is small ($c=4$), it exhibits modest bias when $c = 100$ and the correlation is small ($\rho = 0.1$ or $\rho = 0.5$),  whereas the performance of sPCA methods remains satisfactory. The sPCA methods capture the maximum variance in the data and ultimately yields more favorable results. Although, SDR has modest bias for the case when $c = 100$, it outperforms the existing methods which exhibit substantial bias.  The correlation among the data appears to have no effect on the performance of SDR. However, when $c = 100$, the sPCA methods have improved performance with increasing correlation. This suggest that when variables are strongly correlated, the sPCA methods provide sufficient information for imputation even though the variables screened in SIS may not be identical to the variables in the imputation model. As the dimension of data increases from $p=200$ to $p = 1000$, the results are comparable for all  values of $p$, $c$, and $\rho$. This result suggest that our methods can accommodate different size imputation models. In the case of $c=100$, both our sPCA and SDR proposed methods have improved performance when the covariance structure is block diagonal with compound symmetric blocks (Table 4) as compared to the setting where the covariance matrix is first-order autoregressive (Table 2). Although SDR\_SIR has superior performance within the SDR methods, it is important to note that among the sPCA methods, no method is preferred over the other.

%%%% comparing among the existing approaches
Within the existing methods, in the case where $c=4$, Blasso has better performance in terms of bias, MSE, and coverage rates except in the case $p = 200$ and $\rho = 0.9$. In that exception, DAlasso has better performance. Yet all existing methods underperform our proposed methods when $c=100$. The KNN and RF methods exhibit extreme bias in all scenarios with the exception of KNN\_S with $c=100$, $p=200$, $\rho = 0.5$, and block diagonal covariance structure.  In addition, correlation amongst the data and the number of predictors also appear to have very little influence on the results of existing methods.  In general, the existing methods do not perform well and our proposed approach using dimension reduction techniques yield more favorable results.

\section*{Data example}\label{ex}
We apply the proposed methodology to a prostate cancer study (GEO GDS3289). The data set contains 104 samples of which 34 are benign epithelium samples and 70 nonbenign samples. There were 1,894 fully observed variables which were all used for screening in SIS. In this analysis, we are interested in conducting a logistic regression where we have a binary outcome ($y$) which is 1 if a sample is benign and 0 otherwise. The goal is to test whether a genomic biomarker VPS36 is associated with the outcome. However, VPS36 has 51\% of its values subjected to missingness. For illustration purposes, we also include two fully observed biomarkers as predictors in the logistic regression model. In this analysis, the GS method is not applicable since we do not the underlying true data. In addition, the \textbf{mice} package used to conduct MI in R gives error messages, therefore are not included in our results.

In Table 5, we present the results of our analysis for estimating the parameter ${\theta}_1$, that is the regression parameter associated with VPS36. There were 11 variables screened by SIS for VPS36. The sPCA, SDR, and existing methods give differing results in terms of estimates and p-value. For example, VPS36 is statistically significant using the sPCA methods for dimension reduction before MI but not in SDR. Yet, the direction of the point estimates are the same for all methods except RF. However, RF had extreme bias in the simulations and may be questionable in this application. In addition, the magnitude of the estimates for the sPCA methods are considerably larger but consistent across the four sPCA methods. In contrast, the parameter estimates and p-values are more variable in the three SDR methods. For example, the regression coefficients for VPS36 in the SDR method range from 0.517 to 1.260, with p-values of 0.542 and 0.190, respectively. Our simulations show that performing sPCA before MI generally yields minimal bias and adequate coverage rates, therefore, may be more preferred over the other methods in this application.

\section*{Discussion}\label{disc}
Our work demonstrates the value of dimension reduction techniques in constructing imputation models in the presence of high-dimensional data, particularly when the size of the true imputation model is large. In the settings considered, the proposed methods outperform the existing methods, irregardless of the size of the true imputation model, the number of variables in the data set, and the correlation among the data. In comparing sPCA and SDR to construct imputation models, the sPCA method outperformed SDR in terms of bias, MSE, and coverage rate. A data example using genomics data from a prostate cancer study is used to further illustrate the usefulness of our proposed method.

 We have considered settings under the MAR assumption where a single variable has missing values. In practice, more than one variable of interest may have missing values. Future work can extend our methods to the setting of general missing data patterns with more than one variable missing. In addition, it is of interest to develop methods to handle missing data under the assumption of missing not a random in the presence of high-dimensional data.

	\begin{table}[h]
		\centering
		
		\renewcommand\thetable{1}
		\caption{Simulation results for estimating $\hat{\theta}_1=1$ in the presence of missing data based on 500 Monte Carlo data sets where $n = 100, c = 4, p = 200$ or $p = 1000$ with autoregressive covariance matrix with $\rho$ varying as 0.1, 0.5, and 0.9}
		\resizebox{14.9cm}{4.5cm} {
			%	\begin{adjustbox}{max width=\textwidth,center}
			\begin{tabular}{ p{1.4cm} p{2cm} rrrrp{1.1cm}rrrrp{1.1cm}rrrrrr}
				\hline
				
				& & \multicolumn{5}{l}{$\rho = 0.1$}  & \multicolumn{5}{l}{$\rho = 0.5$} & \multicolumn{5}{l}{$\rho = 0.9$} \\
				\cmidrule(l){3-7} \cmidrule(l){8-12} \cmidrule(l){13-17}
				& Method & Bias & SE & SD & MSE & CR & Bias & SE & SD & MSE & CR & Bias & SE & SD & MSE & CR\\
				\cmidrule(l){1-7} \cmidrule(l){8-12} \cmidrule(l){13-17}
				\\
				&	GS	&	0.001	&	0.086	&	0.088	&	0.008	&	0.944	&	0.000	&	0.081	&	0.080	&	0.006	&	0.962	&	-0.003	&	0.080	&	0.082	&	 0.007	&	0.956	\\
				&	CC	&	-0.172	&	0.103	&	0.107	&	0.041	&	0.610	&	-0.170	&	0.099	&	0.101	&	0.039	&	0.582	&	-0.145	&	0.099	&	0.103	&	 0.032	&	0.696	\\
				&	MI	&	-0.662	&	0.251	&	0.108	&	0.449	&	0.226	&	-0.784	&	0.190	&	0.089	&	0.623	&	0.026	&	-0.778	&	0.209	&	0.085	&	 0.612	&	0.026\\
				\hline
				\\	
				$p = 200$	&	sPCA\_ST	&	-0.026	&	0.100	&	0.097	&	0.010	&	0.940	&	-0.030	&	0.093	&	0.088	&	0.009	&	0.962	&	-0.015	&	 0.094	&	0.097	&	0.010	&	0.956	\\
				&	sPCA\_PMD	&	-0.028	&	0.104	&	0.100	&	0.011	&	0.948	&	-0.027	&	0.097	&	0.091	&	0.009	&	0.970	&	-0.021	&	0.096	&	 0.099	&	0.010	&	0.946	\\
				&	sPCA\_L	&	-0.031	&	0.102	&	0.097	&	0.010	&	0.938	&	-0.035	&	0.095	&	0.088	&	0.009	&	0.958	&	-0.018	&	0.095	&	0.098	 &	0.010	&	0.960	\\
				&	sPCA\_AL	&	-0.030	&	0.101	&	0.097	&	0.010	&	0.938	&	-0.033	&	0.094	&	0.088	&	0.009	&	0.958	&	-0.016	&	0.094	&	 0.098	&	0.010	&	0.960	\\
				
				&	SDR\_SIR	&	-0.028	&	0.106	&	0.107	&	0.012	&	0.940	&	-0.028	&	0.097	&	0.096	&	0.010	&	0.950	&	-0.031	&	0.094	&	 0.094	&	0.010	&	0.932	\\
				&	SDR\_SAVE	&	-0.065	&	0.103	&	0.093	&	0.013	&	0.936	&	-0.067	&	0.098	&	0.088	&	0.012	&	0.900	&	-0.051	&	0.097	&	 0.090	&	0.011	&	0.928	\\
				&	SDR\_PHD	&	-0.060	&	0.104	&	0.097	&	0.013	&	0.936	&	-0.063	&	0.098	&	0.089	&	0.012	&	0.918	&	-0.048	&	0.097	&	 0.091	&	0.011	&	0.930	\\
				&	Blasso	&	-0.059	&	0.113	&	0.098	&	0.013	&	0.944	&	-0.062	&	0.103	&	0.092	&	0.012	&	0.946	&	-0.074	&	0.100	&	0.088	 &	0.013	&	0.908	\\
				&	DAlasso	&	-0.155	&	0.163	&	0.099	&	0.034	&	0.944	&	-0.105	&	0.149	&	0.102	&	0.021	&	0.976	&	-0.049	&	0.126	&	 0.094	&	0.011	&	0.988	\\
				
				& 	KNN\_S &
				-0.254 & 0.150  & 0.143 & 0.085 &  0.640 & -0.281 &  0.146 &  0.141 &  0.099  & 0.524 & -0.289 & 0.142 & 0.126   & 0.100 & 0.468
				
				\\
				&	KNN\_V & -0.313 & 0.158 & 0.126 & 0.114 & 0.504
				& -0.375 &  0.155 & 0.121 &  0.155 & 0.288 & -0.444 & 0.153 &  0.120 & 0.211 & 0.110
				
				\\
				
				& 	RF & -0.320	&	0.176	&	0.119	&	0.116	&	0.624	& 	-0.333	&	0.172	&	0.114	&	0.124	&	0.560	&	-0.305	&	0.160	&	0.107	&	 0.104	&	0.560 \\

				\hline
				\\
				$p = 1000$	&	sPCA\_ST	&	-0.016	&	0.100	&	0.094	&	0.009	&	0.970	&	-0.028	&	0.095	&	0.090	&	0.009	&	0.970	&	-0.015	&	 0.093	&	0.087	&	0.008	&	0.958	\\
				&	sPCA\_PMD	&	-0.020	&	0.101	&	0.094	&	0.009	&	0.966	&	-0.031	&	0.096	&	0.090	&	0.009	&	0.968	&	-0.017	&	0.094	&	 0.088	&	0.008	&	0.956	\\
				&	sPCA\_L	&	-0.023	&	0.102	&	0.094	&	0.009	&	0.964	&	-0.034	&	0.096	&	0.090	&	0.009	&	0.968	&	-0.019	&	0.094	&	0.088	 &	0.008	&	0.956	\\
				&	sPCA\_AL	&	-0.021	&	0.101	&	0.094	&	0.009	&	0.968	&	-0.032	&	0.096	&	0.090	&	0.009	&	0.966	&	-0.017	&	0.094	&	 0.088	&	0.008	&	0.954	\\
				
				&	SDR\_SIR	&	-0.028	&	0.097	&	0.096	&	0.010	&	0.950	&	-0.023	&	0.102	&	0.099	&	0.010	&	0.944	&	-0.034	&	0.097	&	 0.095	&	0.010	&	0.932	\\
				&	SDR\_SAVE	&	-0.067	&	0.098	&	0.088	&	0.012	&	0.900	&	-0.065	&	0.098	&	0.090	&	0.012	&	0.910	&	-0.046	&	0.098	&	 0.091	&	0.010	&	0.932	\\
				&	SDR\_PHD	&	-0.063	&	0.098	&	0.089	&	0.012	&	0.918	&	-0.061	&	0.099	&	0.093	&	0.012	&	0.926	&	-0.043	&	0.098	&	 0.093	&	0.011	&	0.942	\\
				&	Blasso	&	-0.093	&	0.121	&	0.120	&	0.023	&	0.910	&	-0.070	&	0.108	&	0.091	&	0.014	&	0.938	&	-0.079	&	0.102	&	 0.096	&	0.015	&	0.940	\\
				&	DAlasso	&	-0.277	&	0.180	&	0.109	&	0.089	&	0.794	&	-0.250	&	0.176	&	0.107	&	0.074	&	0.846	&	-0.176	&	0.179	&	 0.115	&	0.044	&	0.954	\\
				& KNN\_S & -0.306 & 0.151  & 0.141 & 0.113 &  0.486 &  -0.368 & 0.149  & 0.134 &  0.154 &  0.286 & -0.396 & 0.146 & 0.136  & 0.175 & 0.218 \\
				& KNN\_V & -0.310 & 0.156 &  0.116 & 0.109 & 0.518 & -0.386 & 0.155 & 0.124 & 0.164  & 0.274 &  -0.449 &  0.154  & 0.117 & 0.215 & 0.112\\
				& RF & -0.374	&	0.180	&	0.124	&	0.155	&	0.480	& 	-0.389	&	0.178	&	0.115	&	0.164	&	0.396	&	-0.396	&	0.173	&	0.117	&	 0.171	&	0.364
				\\

				\hline

			\end{tabular}
			%\caption[]{dom}
			%	\end{adjustbox}
		}
	\end{table}

	% % % Table 2
	
	\begin{table}[h]
		%\scriptsize
		\renewcommand\thetable{2}
		\centering
		\caption{Simulation results for estimating $\hat{\theta}_1=1$ in the presence of missing data based on 500 Monte Carlo data sets where $n = 100, c = 100, p = 200$ or $p = 1000$ with autoregressive covariance matrix with $\rho$ varying as 0.1, 0.5, and 0.9 }
		\resizebox{14.9cm}{4.5cm} {
			%	\begin{adjustbox}{max width=\textwidth,center}
			\begin{tabular}{ p{1.4cm} p{2cm} rrrrp{1.1cm}rrrrp{1.1cm}rrrrrr}
				\hline

				& & \multicolumn{5}{l}{$\rho = 0.1$}  & \multicolumn{5}{l}{$\rho = 0.5$} & \multicolumn{5}{l}{$\rho = 0.9$} \\
				\cmidrule(l){3-7} \cmidrule(l){8-12} \cmidrule(l){13-17}
				& Method & Bias & SE & SD & MSE & CR & Bias & SE & SD & MSE & CR & Bias & SE & SD & MSE & CR\\
				\cmidrule(l){1-7} \cmidrule(l){8-12} \cmidrule(l){13-17}
				\\
				&	GS	&	0.011	&	0.156	&	0.158	&	0.025	&	0.940	&	-0.010	&	0.136	&	0.139	&	0.019	&	0.946	&	0.001	&	0.086	&	0.090	&	 0.008	&	0.948	\\
				&	CC	&	-0.334	&	0.168	&	0.172	&	0.141	&	0.518	&	-0.322	&	0.150	&	0.158	&	0.128	&	0.414	&	-0.233	&	0.112	&	0.113	&	 0.067	&	0.484	\\
				&	MI	&	-0.800	&	0.209	&	0.094	&	0.649	&	0.038	&	-0.772	&	0.216	&	0.089	&	0.604	&	0.040	&	-0.672	&	0.265	&	0.099	&	 0.461	&	0.232	\\
				\hline
				\\
				$p = 200$	&	sPCA\_ST	&	-0.098	&	0.212	&	0.203	&	0.050	&	0.964	&	-0.067	&	0.174	&	0.162	&	0.031	&	0.972	&	-0.013	&	 0.108	&	0.111	&	0.013	&	0.962	\\
				&	sPCA\_PMD	&	-0.105	&	0.214	&	0.204	&	0.053	&	0.966	&	-0.075	&	0.177	&	0.164	&	0.033	&	0.972	&	-0.016	&	0.110	&	 0.112	&	0.013	&	0.962	\\
				&	sPCA\_L	&	-0.111	&	0.217	&	0.203	&	0.053	&	0.962	&	-0.085	&	0.180	&	0.166	&	0.035	&	0.972	&	-0.022	&	0.112	&	0.114	 &	0.013	&	0.964	\\
				&	sPCA\_AL	&	-0.108	&	0.216	&	0.203	&	0.053	&	0.966	&	-0.079	&	0.179	&	0.165	&	0.034	&	0.972	&	-0.017	&	0.110	&	 0.113	&	0.013	&	0.966	\\
				
				&	SDR\_SIR	&	-0.311	&	0.229	&	0.232	&	0.151	&	0.748	&	-0.242	&	0.200	&	0.189	&	0.094	&	0.820	&	-0.062	&	0.117	&	 0.118	&	0.018	&	0.928	\\
				&	SDR\_SAVE	&	-0.153	&	0.217	&	0.209	&	0.067	&	0.956	&	-0.128	&	0.186	&	0.162	&	0.043	&	0.962	&	-0.077	&	0.119	&	 0.108	&	0.018	&	0.938	\\
				&	SDR\_PHD	&	-0.188	&	0.224	&	0.218	&	0.083	&	0.920	&	-0.151	&	0.192	&	0.171	&	0.052	&	0.946	&	-0.071	&	0.120	&	 0.112	&	0.018	&	0.936	\\
				
				&	Blasso	&	-0.450	&	0.238	&	0.143	&	0.223	&	0.580	&	-0.420	&	0.219	&	0.122	&	0.191	&	0.546	&	-0.094	&	0.129	&	 0.109	&	0.021	&	0.928	\\
				&	DAlasso	&	-0.466	&	0.253	&	0.146	&	0.239	&	0.608	&	-0.393	&	0.231	&	0.130	&	0.171	&	0.682	&	-0.063	&	0.139	&	 0.116	&	0.017	&	0.970	\\
				& KNN\_S & -0.294 & 0.222 & 0.216 & 0.133 & 0.750 & -0.279 & 0.205 & 0.193 &  0.115 & 0.738 & -0.035 &  0.154 & 0.133 &  0.019 & 0.958  \\
				& KNN\_V  & -0.289 & 0.233 & 0.189 & 0.119 & 0.810 & -0.310 & 0.217 & 0.161 & 0.122 &  0.760 & -0.277 & 0.182 &  0.127 &  0.093 & 0.754\\
				& RF & -0.442	&	0.250	&	0.147	&	0.217	&	0.664	& 	-0.434	&	0.232	&	0.136	&	0.207	&	0.566	&	-0.246	&	0.182	&	0.117	&	 0.074	&	0.836 \\

				\hline
				\\
				$p = 1000$	&	sPCA\_ST	&	-0.080	&	0.210	&	0.195	&	0.044	&	0.982	&	-0.065	&	0.175	&	0.165	&	0.031	&	0.978	&	-0.022	&	 0.108	&	0.104	&	0.011	&	0.964	\\
				&	sPCA\_PMD	&	-0.085	&	0.212	&	0.195	&	0.045	&	0.978	&	-0.070	&	0.177	&	0.168	&	0.033	&	0.976	&	-0.025	&	0.110	&	 0.105	&	0.012	&	0.964	\\
				&	sPCA\_L	&	-0.099	&	0.217	&	0.193	&	0.047	&	0.974	&	-0.084	&	0.182	&	0.168	&	0.035	&	0.980	&	-0.032	&	0.112	&	0.106	 &	0.012	&	0.964	\\
				&	sPCA\_AL	&	-0.094	&	0.215	&	0.194	&	0.046	&	0.974	&	-0.079	&	0.180	&	0.168	&	0.034	&	0.980	&	-0.027	&	0.110	&	 0.105	&	0.012	&	0.964	\\
				
				&	SDR\_SIR	&	-0.350	&	0.229	&	0.217	&	0.170	&	0.688	&	-0.314	&	0.209	&	0.201	&	0.139	&	0.702	&	-0.079	&	0.124	&	 0.116	&	0.020	&	0.930	\\
				&	SDR\_SAVE	&	-0.135	&	0.215	&	0.192	&	0.055	&	0.956	&	-0.122	&	0.187	&	0.173	&	0.045	&	0.948	&	-0.078	&	0.119	&	 0.105	&	0.017	&	0.944	\\
				&	SDR\_PHD	&	-0.183	&	0.225	&	0.199	&	0.073	&	0.926	&	-0.158	&	0.198	&	0.182	&	0.058	&	0.926	&	-0.076	&	0.122	&	 0.106	&	0.017	&	0.952	\\
				
				&	Blasso	&	-0.483	&	0.245	&	0.141	&	0.253	&	0.542	&	-0.483	&	0.232	&	0.131	&	0.250	&	0.456	&	-0.213	&	0.158	&	0.102	 &	0.056	&	0.844	\\
				
				&	DAlasso	&	-0.447	&	0.257	&	0.148	&	0.222	&	0.670	&	-0.408	&	0.240	&	0.139	&	0.185	&	0.700	&	-0.167	&	0.185	&	 0.121	&	0.042	&	0.954	\\
				& KNN\_S &  -0.339 & 0.222 & 0.204 & 0.156 & 0.712 & -0.320 & 0.207 & 0.195  & 0.140 &  0.688 & -0.185 & 0.165  & 0.142 & 0.055 & 0.830 \\
				& KNN\_V & -0.303 &  0.231 &  0.177 & 0.123 & 0.812 & -0.302 & 0.217 & 0.169 & 0.119 &  0.770 &  -0.295 &  0.181 &  0.124 & 0.102 & 0.664
				\\
				
				& RF & -0.470	&	0.250	&	0.136	&	0.240	&	0.608	& 	-0.478	&	0.234	&	0.141	&	0.249	&	0.470	&	-0.357	&	0.195	&	0.111	&	 0.140	&	0.636
				\\	
				\hline
				
			\end{tabular}
		}
	\end{table}

	% % % Table 3
	
	\begin{table}[h]
		%\scriptsize
		%\renewcommand\thetable{3}
		\centering
		\caption{Simulation results for estimating $\hat{\theta}_1=1$ in the presence of missing data based on 500 Monte Carlo data sets where $n = 100, c = 4, p = 200$ or $p = 1000$ with block diagonal matrix with compound symmetric blocks and $\rho$ varying as 0.1, 0.5, and 0.9 }
		\resizebox{14.9cm}{4.5cm} {
			%	\begin{adjustbox}{max width=\textwidth,center}
			\begin{tabular}{ p{1.4cm} p{2cm} rrrrp{1.1cm}rrrrp{1.1cm}rrrrrr}
				\hline
				& & \multicolumn{5}{l}{$\rho = 0.1$}  & \multicolumn{5}{l}{$\rho = 0.5$} & \multicolumn{5}{l}{$\rho = 0.9$} \\
				\cmidrule(l){3-7} \cmidrule(l){8-12} \cmidrule(l){13-17}
				& Method & Bias & SE & SD & MSE & CR & Bias & SE & SD & MSE & CR & Bias & SE & SD & MSE & CR\\
				\cmidrule(l){1-7} \cmidrule(l){8-12} \cmidrule(l){13-17}
				\\
				&	GS	&	 -0.004 & 0.066 & 0.066 & 0.004 & 0.956 & 0.003 & 0.061 & 0.062 & 0.004 &0.942 &
				0.003 & 0.059 &  0.058 & 0.003 & 0.948\\

				&	CC	&  -0.129  & 0.086 & 0.090 & 0.025  & 0.686 & -0.105  & 0.082 & 0.079 & 0.017  & 0.756 & -0.090  & 0.080 & 0.084 & 0.015  & 0.790 \\
				
				&	MI	& -0.742 & 0.205 & 0.102 & 0.561 & 0.070 &
				-0.670 & 0.248 & 0.098 & 0.458 & 0.182 &
				-0.642 & 0.278 & 0.089 & 0.421 & 0.326
						\\
				\hline
				\\
				$p = 200$	&	sPCA\_ST	&  -0.034 & 0.088 & 0.084 & 0.008  & 0.954 & -0.014 & 0.087 & 0.081 &  0.007  & 0.966 &
				0.002 & 0.083 & 0.080 & 0.006 & 0.970
				\\
				&	sPCA\_PMD	&	 -0.029 & 0.092 & 0.086 & 0.008 & 0.958
				& -0.016  & 0.087 & 0.079 & 0.007 & 0.966 &
				-0.004 & 0.080 & 0.078  & 0.006 & 0.958
				\\
				&	sPCA\_L	&	 -0.037 & 0.089 & 0.084  & 0.008  & 0.956 & -0.015 & 0.087 & 0.081 &  0.007 &  0.966 & -0.003 & 0.086 & 0.082 & 0.007 & 0.968
				\\
				&	sPCA\_AL	&	 -0.036 & 0.089 &  0.084 & 0.008 & 0.956
				& -0.014 & 0.087 & 0.081 & 0.007 & 0.966
				& 0.001 & 0.084 & 0.081 & 0.006 & 0.972
				\\
				
				&	SDR\_SIR	&	0.002 & 0.081 & 0.075 & 0.006 & 0.972 & -0.015 & 0.074 & 0.073 & 0.006 & 0.954 &  -0.006  & 0.082  & 0.086 & 0.007  & 0.956
				
				\\
				&	SDR\_SAVE	&	 -0.059 & 0.091 & 0.076 & 0.009 & 0.940
				& -0.043 & 0.087 & 0.077 & 0.008 & 0.956
				& -0.073 & 0.092  & 0.086 & 0.013 & 0.902
				\\
				&	SDR\_PHD	&	 -0.052 & 0.091 & 0.076  & 0.009 &  0.944 &
				-0.041 & 0.086 & 0.075  & 0.007 & 0.958
				& -0.066 &  0.091 & 0.086 & 0.012  & 0.926
				\\
				&	Blasso	&  0.042 &  0.096 & 0.089 & 0.010 & 0.948 &
					0.040 & 0.091 & 0.091 &  0.010 & 0.942 &
					 0.028 & 0.091 & 0.083 &  0.008 & 0.968
					 	\\
				&	DAlasso	&	 -0.185 & 0.169 & 0.098  & 0.044 & 0.938 & -0.108 & 0.147 & 0.096 & 0.021  & 0.980 &
				-0.005 & 0.109 & 0.080 & 0.006 & 0.992
				
					\\
				& KNN\_S &  -0.187 & 0.137 & 0.126 & 0.051 & 0.742
			& -0.101 & 0.127 & 0.110 & 0.022 & 0.908 &
			 -0.055 & 0.119 &  0.105 & 0.014 & 0.952
				\\
				& KNN\_V &  -0.256 & 0.147 & 0.109 & 0.077 & 0.644
				& -0.307 & 0.145 & 0.116 & 0.108 & 0.454
				 & -0.375 & 0.144 & 0.124 & 0.156 & 0.234
				 \\
				& RF & -0.288 & 0.163 & 0.106 &  0.094 & 0.666
			&	-0.422 & 0.148 & 0.096 & 0.187 & 0.126&
				-0.145 & 0.128 & 0.087 & 0.028  & 0.896
				
				\\
				\hline
				\\
				$p = 1000$	&	sPCA\_ST & 0.015 & 0.082 & 0.084 & 0.007 & 0.946 & 0.012 & 0.083 & 0.084 & 0.007 & 0.960 & 0.019 & 0.119 & 0.119 & 0.015 & 0.952\\

				&	sPCA\_PMD	&	 0.013 & 0.082 & 0.083 & 0.007 & 0.952 & 0.003 & 0.087 & 0.086 & 0.007 & 0.956 & -0.002 & 0.120 & 0.118 & 0.014 & 0.946
				\\
				&	sPCA\_L	&	 0.013 & 0.083 & 0.083 & 0.007 & 0.946 & 0.013 & 0.083 & 0.084 & 0.007 & 0.958 & 0.015 & 0.118 & 0.118 & 0.014 & 0.950 \\
				&	sPCA\_AL	&	 0.014 & 0.082 & 0.083 & 0.007 & 0.946 & 0.013 & 0.083 & 0.084 & 0.007 & 0.958 & 0.018 & 0.119 & 0.119 & 0.014 & 0.952
				
				\\

				&	SDR\_SIR	&	 0.001 & 0.081 & 0.083 & 0.007 & 0.944 &  -0.019  & 0.085 & 0.086  & 0.008  & 0.934 & -0.116 & 0.141 & 0.157 & 0.038 & 0.874

				\\
				&	SDR\_SAVE	&	-0.062 & 0.087 & 0.079 & 0.010 & 0.924
				& -0.029 & 0.087 & 0.080 & 0.007  & 0.948
				& -0.093  & 0.118 & 0.109 & 0.021 & 0.898
				\\
				&	SDR\_PHD	&	 -0.057 & 0.087 & 0.081 & 0.010 & 0.934
				& -0.026 & 0.087 & 0.081  & 0.007 & 0.954
				& -0.091 & 0.120 & 0.111 &  0.021 & 0.904
				\\
				&	Blasso	&	0.031 & 0.095 & 0.091 & 0.009 & 0.960 & -0.118 & 0.159 & 0.196  & 0.052 & 0.926
			&	 -0.735 & 0.135 & 0.167 & 0.567  & 0.056
					\\
				
				&	DAlasso	& -0.346 & 0.171 & 0.087 & 0.127 & 0.506 & -0.433 & 0.168 &  0.102 &  0.198 & 0.200
				& -0.528 & 0.182 & 0.106 & 0.290 & 0.098
				
					\\
				
				& KNN\_S & -0.242 & 0.137  & 0.124 & 0.074 & 0.602
				& -0.359 & 0.129 & 0.122 &  0.143 & 0.174 &
				 -0.819 & 0.110 & 0.105 &  0.681 &  0.000
				 \\
				& KNN\_V &  -0.298 & 0.147  & 0.112 & 0.102 &  0.466 &
				-0.616  & 0.131 &  0.118  & 0.394  & 0.004 &
				-0.953 &  0.096 & 0.092 & 0.916 & 0.000
				 \\
				
				& RF & -0.356 & 0.167  & 0.105 &  0.138 & 0.422
				& -0.235 & 0.151  & 0.101 &  0.065 &  0.748 &
			-0.719  & 0.129 &  0.102 & 0.528 & 0.000
				\\
				
				\hline
				
			\end{tabular}
		}
	\end{table}

	% % % Table 4
	
	\begin{table}[h]
		%\scriptsize
		%\renewcommand\thetable{3}
		\centering
		\caption{Simulation results for estimating $\hat{\theta}_1=1$ in the presence of missing data based on 500 Monte Carlo data sets where $n = 100, c = 100, p = 200$ or $p = 1000$ with block diagonal matrix with compound symmetric blocks and $\rho$ varying as 0.1, 0.5, and 0.9 }
		\resizebox{14.9cm}{4.5cm} {
			%	\begin{adjustbox}{max width=\textwidth,center}
			\begin{tabular}{ p{1.4cm} p{2cm} rrrrp{1.1cm}rrrrp{1.1cm}rrrrrr}
				\hline
				& & \multicolumn{5}{l}{$\rho = 0.1$}  & \multicolumn{5}{l}{$\rho = 0.5$} & \multicolumn{5}{l}{$\rho = 0.9$} \\
				\cmidrule(l){3-7} \cmidrule(l){8-12} \cmidrule(l){13-17}
				& Method & Bias & SE & SD & MSE & CR & Bias & SE & SD & MSE & CR & Bias & SE & SD & MSE & CR\\
				\cmidrule(l){1-7} \cmidrule(l){8-12} \cmidrule(l){13-17}
				\\
				&	GS	&	0.001	&	0.107	&	0.115	&	0.013	&	0.954	&	0.002	&	0.073	&	0.074	&	0.005	&	0.954	&	-0.004	&	0.064	&	0.063	&	 0.004	&	0.954	\\
				&	CC	&	-0.227	&	0.131	&	0.133	&	0.069	&	0.596	&	-0.150	&	0.101	&	0.101	&	0.033	&	0.724	&	-0.115	&	0.092	&	0.094	&	 0.022	&	0.772	\\
				&	MI	&	-0.778	&	0.192	&	0.092	&	0.613	&	0.026	&	-0.502	&	0.301	&	0.109	&	0.264	&	0.770	&	-0.585	&	0.300	&	0.091	&	 0.351	&	0.564	\\
				\hline
				\\
				$p = 200$	&	sPCA\_ST	&	-0.047	&	0.136	&	0.130	&	0.019	&	0.974	&	-0.002	&	0.100	&	0.097	&	0.009	&	0.950	&	-0.006	&	 0.096	&	0.095	&	0.009	&	0.966	\\
				&	sPCA\_PMD	&	-0.052	&	0.138	&	0.130	&	0.020	&	0.968	&	-0.009	&	0.102	&	0.097	&	0.009	&	0.950	&	-0.007	&	0.097	&	 0.095	&	0.009	&	0.964	\\
				&	sPCA\_L	&	-0.063	&	0.141	&	0.132	&	0.021	&	0.966	&	-0.009	&	0.102	&	0.098	&	0.010	&	0.952	&	-0.010	&	0.098	&	0.095	 &	0.009	&	0.966	\\
				&	sPCA\_AL	&	-0.055	&	0.139	&	0.130	&	0.020	&	0.968	&	-0.005	&	0.101	&	0.098	&	0.010	&	0.950	&	-0.007	&	0.097	&	 0.095	&	0.009	&	0.966	\\
				
				&	SDR\_SIR	&	-0.171	&	0.164	&	0.147	&	0.051	&	0.852	&	-0.027	&	0.107	&	0.106	&	0.012	&	0.940	&	-0.016	&	0.088	&	 0.098	&	0.010	&	0.938	\\
				&	SDR\_SAVE	&	-0.123	&	0.151	&	0.120	&	0.030	&	0.942	&	-0.060	&	0.107	&	0.097	&	0.013	&	0.928	&	-0.027	&	0.094	&	 0.093	&	0.009	&	0.954	\\
				&	SDR\_PHD	&	-0.138	&	0.157	&	0.129	&	0.035	&	0.924	&	-0.056	&	0.109	&	0.099	&	0.013	&	0.940	&	-0.026	&	0.093	&	 0.093	&	0.009	&	0.952	\\
				&	Blasso	&	-0.251	&	0.187	&	0.128	&	0.079	&	0.830	&	-0.020	&	0.125	&	0.103	&	0.011	&	0.988	&	0.035	&	0.101	&	 0.089	&	0.009	&	0.966	\\
				&	DAlasso	&	-0.284	&	0.200	&	0.123	&	0.096	&	0.826	&	-0.066	&	0.144	&	0.107	&	0.016	&	0.990	&	-0.039	&	0.133	&	 0.097	&	0.011	&	0.990	\\
				& KNN\_S & -0.122 & 0.174  & 0.155  & 0.039 &  0.930 & 0.008 & 0.138 & 0.117 & 0.014 & 0.982 & -0.051 & 0.128 & 0.112 & 0.015  & 0.962\\
				& KNN\_V &  -0.232 & 0.191  & 0.153 & 0.077 & 0.806 & -0.306 &  0.169 & 0.127 & 0.110 & 0.596 & -0.401 & 0.157 &  0.135 &  0.179 & 0.248\\
				& RF & -0.370	&	0.204	&	0.130	&	0.154	&	0.622	& 	-0.269	&	0.170	&	0.099	&	0.082	&	0.758	&	-0.242	&	0.153	&	0.102	&	 0.069	&	0.714
				\\
				\hline
				\\
				$p = 1000$	&	sPCA\_ST	&	0.010	&	0.099	&	0.106	&	0.011	&	0.934	&	0.055	&	0.074	&	0.079	&	0.009	&	0.870	&	0.014	&	 0.102	&	0.096	&	0.009	&	0.954	\\
				&	sPCA\_PMD	&	0.004	&	0.100	&	0.106	&	0.011	&	0.936	&	0.044	&	0.076	&	0.080	&	0.008	&	0.906	&	-0.007	&	0.102	&	 0.095	&	0.009	&	0.974	\\
				&	sPCA\_L	&	0.000	&	0.101	&	0.106	&	0.011	&	0.944	&	0.053	&	0.074	&	0.080	&	0.009	&	0.882	&	0.010	&	0.102	&	0.096	 &	0.009	&	0.956	\\
				&	sPCA\_AL	&	0.006	&	0.100	&	0.106	&	0.011	&	0.934	&	0.055	&	0.074	&	0.080	&	0.009	&	0.872	&	0.013	&	0.102	&	 0.096	&	0.009	&	0.956	\\

				&	SDR\_SIR	&	-0.068	&	0.130	&	0.125	&	0.020	&	0.934	&	0.001	&	0.084	&	0.085	&	0.007	&	0.936	&	-0.083	&	0.121	&	 0.130	&	0.024	&	0.898	\\
				&	SDR\_SAVE	&	-0.076	&	0.114	&	0.102	&	0.016	&	0.936	&	-0.024	&	0.085	&	0.079	&	0.007	&	0.968	&	-0.095	&	0.108	&	 0.099	&	0.019	&	0.902	\\
				&	SDR\_PHD	&	-0.075	&	0.119	&	0.109	&	0.017	&	0.944	&	-0.020	&	0.085	&	0.079	&	0.007	&	0.966	&	-0.095	&	0.110	&	 0.101	&	0.019	&	0.908	\\
				&	Blasso	&	-0.170	&	0.164	&	0.104	&	0.040	&	0.920	&	-0.086	&	0.139	&	0.146	&	0.029	&	0.972	&	-0.623	&	0.139	&	0.218	 &	0.435	&	0.148	\\
				
				&	DAlasso	&	-0.192	&	0.195	&	0.118	&	0.051	&	0.948	&	-0.209	&	0.183	&	0.141	&	0.063	&	0.922	&	-0.476	&	0.209	&	0.149	 &	0.249	&	0.414	\\
				
				& KNN\_S & -0.028  & 0.147 &  0.129 & 0.017 & 0.974 & -0.206 & 0.130 & 0.114 & 0.055 & 0.654 & -0.832 & 0.101 & 0.103 & 0.703 & 0.000\\
				& KNN\_V & -0.240 & 0.172 & 0.126 & 0.073 & 0.778 & -0.575 & 0.140 &  0.120 & 0.345 & 0.004 &
				-0.954 & 0.091 & 0.091 & 0.919 & 0.000\\
				
				& RF & -0.347	&	0.186	&	0.109	&	0.132	&	0.616	& 	-0.342	&	0.153	&	0.097	&	0.126	&	0.396	&	-0.738	&	0.116	&	0.092	&	 0.553	&	0.000 \\
				
				\hline
				
			\end{tabular}
		}
	\end{table}

	\begin{table}[h]
		%\scriptsize
		
		\centering
		\caption{Estimation of the predictor variable $\theta_1$ that is associated with the incomplete biomarker VPS36 in logistic regression using complete case analysis (CC), four sparse PCA methods (sPCA), three SDR methods, Bayesian Lasso (Blasso), direct use of adaptive lasso (DAlasso), and random forest multiple imputation (RF)  using the prostate cancer study }
		%\resizebox{17cm}{!} {
		\begin{tabular}{lccc}
			\hline
			
				Method & Estimate & SE & p-value \\
			\hline 
			CC & 1.336 & 1.165 & 0.2515  \\

			Blasso & 0.226 & 0.692 & 0.7463  \\
			DAlasso & 0.908 & 0.985 & 0.3631  \\
			 RF & -2.922 & 2.002 & 0.1550  \\
			\hline
			 sPCA\_ST & 2.290 & 0.942 & 0.0183  \\
	    	sPCA\_PMD & 2.276 & 0.945 & 0.0192 \\
			 sPCA\_L & 2.170 & 0.925 & 0.0228  \\
			 sPCA\_AL & 2.256 & 0.940 & 0.0199 \\
			
			SDR\_SIR & 1.260 & 0.950 & 0.1898 \\
			 SDR\_SAVE & 0.864 & 0.813 & 0.2928 \\
			SDR\_PHD & 0.517 & 0.842  & 0.5418 \\

			\hline
		\end{tabular}

	\end{table}

\end{document}